\documentstyle[epsfig]{mn2e}

\title[Hydrostatic equilibrium of neutron star models]{Hydrostatic equilibrium of causally consistent and dynamically stable neutron star models}

\author[P. S. Negi]{P. S. Negi$^{1}$\thanks{E-mail:
negi@aries.ernet.in; psnegi\_nainital@yahoo.com} \\
$^{1}$ Department of Physics, Kumaun University,
              Nainital - 263 002, India \\
}
\date{Accepted ------ .
      Received ------ ;
      }
\pagerange{\pageref{firstpage}--\pageref{lastpage}} 
\pubyear{2006}

\begin{document}

\maketitle

\label{firstpage}

\begin{abstract}
  We show that the mass-radius $(M-R)$ relation corresponding to the stiffest equation of state (EOS) does not provide the necessary and sufficient condition of dynamical stability for the equilibrium configurations, since such configurations can not satisfy the `compatibility criterion'. In this connection, we construct sequences composed of core-envelope models such that, like the stiffest EOS, each member of these sequences satisfy the extreme case of causality condition, $v = c = 1$, at the centre. We, thereafter, show that the $M-R$ relation corresponding to the said core-envelope model sequences can provide the necessary and sufficient condition of dynamical stability only when the `compatibility criterion' for these sequences is `appropriately' satisfied. However, the fulfillment of `compatibility criterion' can remain satisfied even when the $M-R$ relation does not provide the necessary and sufficient condition of dynamical stability for the equilibrium configurations.
 In continuation to the results of previous study, these results explicitly show that the `compatibility criterion' {\em independently} provides, in general, the {\em necessary} and {\em sufficient} condition of hydrostatic equilibrium for any regular sequence. Beside its fundamental feature, this study can also explain simultaneously, both (the higher as well as lower) values of the glitch healing parameter observed for the Crab and the Vela-like pulsars respectively, on the basis of starquake model of glitch generation.

\end{abstract}


\section{Introduction}

        Characteristics of the super-dense objects  like  neutron 
stars (NSs) are based on the  calculations  of  the EOS for the  matter  at  very  high  densities.  However,  the 
nuclear interactions beyond the  density  of $  \sim 10^{14} {\rm g\, cm}^{-3}$ 
are 
empirically not well known (Dolan 1992) and  the all known  EOSs 
are only extrapolations of the empirical results far beyond  this 
density range. In this regard, various EOSs based  on  theoretical 
manipulations are available in the literature (Arnett  \&  Bowers 
1977). Since the status  of  EOS  of  nuclear   matter   cannot   be   established 
empirically beyond a certain density range, one can apply  physical  constraints  to  obtain  an upper bound of neutron star mass.  Brecher  and  Caporaso  (1976) 
assumed that the speed of sound in the nuclear matter equals  the 
speed of light and obtained a value of $4.8 M_\odot$  as an  upper  limit 
of the neutron star masses. However, the matter described by this stiffest EOS, $(dP/dE) = 1$ (geometrized  units)
or $P = (E - E_s)$ [where $P$ is the pressure, $E$ is the energy-density and $E_s   =  2 \times 10^{14} {\rm g\, cm}^{-3}$ represents the  surface 
density] has a super-dense self-bound state at  the  surface  of 
the configuration where the pressure is vanishingly small,  which 
represent the `abnormal state of matter' (Lee  1975;  Haensel  \& 
Zdunik 1989). This `abnormality' may be specified as -  (i)  the 
pressure vanishes at the average nuclear densities, and (ii)  the 
speed of sound is equal to that of light even when  the  pressure 
is vanishing small. This `abnormality'  can  be  removed  if  we
ensure  continuity  of pressure, density and both of the  metric  parameters at the boundary of the structure (Negi \& Durgapal 2000).
 
     Earlier, Rhoades and Ruffini (1974), without going into  the 
details of  the  nuclear  interactions,  assumed  that  beyond  a 
certain density $ 4.6 \times 10^{14} {\rm g\, cm}^{-3}$  (the range of densities  where 
no extrapolated EOS is known) the EOS in the core is given by the 
criterion that the speed of sound attains  the  speed  of  light, 
that is, $(dP/dE)= 1$, and matched the core to an envelope with the 
BPS (Baym, Pethick \& Sutherland 1971) EOS and obtained an  upper 
limit for  the  neutron  star  mass  as  $3.2 M_\odot$ .  Hartle  (1978) 
emphasized that the maximum masses of neutron stars  obtained  in 
this manner involve a scale factor, $k  =  [E_m /10^{14} {\rm g\, cm}^{-3}]^{-1/2}$, 
such that, the matching density, $E_m$ , plays a  sensitive  role  to 
obtain an upper bound on neutron star masses. Usually, $k$ is taken 
to be equal to or  greater  than  one  in  all  the  conventional 
models. For the densities less than $E_m$ , the matter composing  the 
object is assumed to be known and unique. That is, the EOS of the 
envelope of these stars are chosen so that the `abnormalities' in 
the sense mentioned above are removed. Friedman and Ipser  (1987) 
calculated the masses of the neutron stars for  different  values 
of the matching densities using EOS given by BPS and NV (Negele \& 
Vautherin 1973),  respectively,  and  concluded  that   the  EOS 
chosen for the envelope does not make any significant  difference 
in the results, because in each case the  mass  in  the  envelope 
turns out to be insignificant as  compared  to  that  of  the  core 
containing the most  stiff  material. 

In order to implement this `insignificant' mass of the envelope component, Crawford \& Demia\'{n}ski (2003) have recently constructed the NS models for seven representative EOSs of dense
 nuclear matter by covering a range of NS masses. They have computed the `fractional moment of inertia' of the core component which is defined as the glitch healing parameter, $Q$, in the starquake mechanism of glitch generation as

\begin{equation}
Q = \frac {I_{\rm core}}{I_{\rm total}},
\end{equation}
where $I_{\rm total}$ represents the moment of inertia of the entire configuration.
Their study shows that the much larger values of $Q(\geq 0.7)$ for the Crab pulsar are fulfilled by all but 
  the six EOSs considered in the study corresponding to a `realistic' neutron star mass range 
$1.4\pm 0.2M_\odot$. On the other hand, for the much lower values of $Q(\leq 0.2)$ corresponding to the case of the Vela pulsar, their models predict a NS mass $\leq 0.5 M_\odot$ which is too low as compared to the `realistic' NS mass range. Thus, their study concludes the `starquake' as a feasible mechanism for glitch generation in the Crab-like pulsars (corresponding to a larger value of $Q$) and the `vortex unpinning', the another mechanism of glitch generation, suitable for the Vela-like pulsars (corresponding to a lower values of $Q$). However, it seems quite surprising that if the internal structure of NSs is supposed to be described by the same two-component conventional model, then why different kinds of glitch mechanisms are required to explain a glitch!

This is the purpose of this study to provide an insight about a fundamental `theorem' concerning the hydrostatic equilibrium of NS models (which was laking in the earlier studies) and to show that as soon as this theorem is implemented to an appropriate conventional NS sequence, the various shortcomings of the conventional NS models (as mentioned above) can be resolved.


\section{Removal of abnormalities from the stiffest EOS and 
 boundary conditions for neutron star models compatible with hydrostatic equilibrium}

     In order to construct such an appropriate sequence of NS models, consistent with causality and dynamical stability, we offer here an entirely different approach to the whole 
problem  which will not only remove the `abnormalities'  of  the stiffest EOS (as  discussed 
earlier in detail under section 1) but can also assure the {\em necessary and sufficient} condition of hydrostatic equilibrium for the resulting configuration. Since, as we would show later in section 3 that the $M-R$ relations corresponding to the configurations (i) governed by the pure stiffest EOS,
and (ii) those resulting from the removal of `abnormalities' from the stiffest EOS (core-envelope models), do not provide the necessary and sufficient condition of dynamical stability unless the `compatibility criterion' (Negi \& Durgapal 2001) which states that: ``for each and every assigned value of $\sigma [\equiv (P_0/E_0) \equiv $ the ratio  of central  pressure  to   central   energy-density], the compactness ratio $u (\equiv M/R)$ of the entire configuration should not  exceed the compactness ratio, $u_h$, of the corresponding homogeneous density sphere (that is, $u \leq  u_h$)'' is `appropriately' satisfied by such configurations.

     The present approach is based on the `theorem' applicable to a wide range of conventional NS sequences (including the core-envelope models) provided that every member of such sequences satisfies the condition $(dP/dE)_0 = 1$ (here and elsewhere in the paper, the subscript `0' represents the value of the corresponding quantity at the centre of the configuration).  This theorem asserts that in order to assure the necessary and
sufficient condition of dynamical stability for the mass, the maximum stable value of $u$
(corresponding to
the case of first maxima among masses in the mass-radius ($M-R$) relation) and the corresponding central value
of `local' adiabatic index $(\Gamma_1)_0 (\equiv [(P_0 + E_0)/P_0](dP/dE)_0)$ of such sequences must satisfy the inequalities, $u_{\rm max} \leq u_{\rm max,abs} \cong 0.3406$
and $(\Gamma_1)_0 \leq (\Gamma_1)_{0,\rm max, abs} \cong 2.5946$ respectively (Negi 2007). In addition to the result of previous study regarding the `compatibility criterion', we would show in the present study that ``the $M-R$ relation
(or the mass-central density relation) corresponding to an equilibrium sequence provides the necessary and sufficient condition of dynamical stability only when the `compatibility criterion' for the equilibrium sequence is also satisfied. But the fulfillment of the `compatibility criterion' does not depend upon the fulfillment of the necessary and sufficient condition of dynamical stability provided by the $M-R$ relation''. The proof of this statement would explicitly show that the fulfillment of `compatibility criterion' independently provides, in general, the necessary and sufficient condition of hydrostatic equilibrium for any sequence composed of regular configurations, since the first sentence of this statement has already been proved in the previous study for the equilibrium sequences of the type mentioned above (Negi 2007).

In order to provide a proof of the last statement, the present paper deals with the construction of a four step consecutive study in the following manner

{\bf Step (1):} We would construct the NS models corresponding to the pure stiffest EOS, $dP/dE = 1$, and show that the $M-R$ relation corresponding to this sequence does not provide the necessary and sufficient condition of dynamical stability, since the `compatibility criterion' can not be fulfilled by the equilibrium configurations corresponding to this `abnormal' EOS.

{\bf Step (2):} We would construct the NS sequences composed of core-envelope models  by considering the stiffest EOS, $dP/dE = 1$, in the core and the EOS of classical polytrope, d${\rm ln}P/$d${\rm ln}\rho = \Gamma_1$ (where $\rho$ is the density of the rest-mass and $\Gamma_1$ is the constant adiabatic index; see, e.g. Tooper 1965), in the envelope (so that each member of these sequences satisfy the extreme case of causality condition, $v = c = 1$, at the centre) for an `arbitrarily' assigned value of pressure to density ratio, $(P_b/E_b) = (P_b/E_b)_1$ (say), at the core-envelope boundary. Though this procedure removes the `abnormalities' from the stiffest EOS, but as we would show later in sec.3 that the $M-R$ relation corresponding to the NS models with an envelope $\Gamma_1 = 5/3$ and 2, do not provide the necessary and sufficient condition of dynamical stability, since  the `compatibility criterion' remains  unsatisfied by various stable (as well as unstable)  models corresponding to the said sequences.

{\bf Step (3):} In this sub-section we would re-construct the NS sequence composed of core-envelope models as described in step (2) for the same boundary value of $(P_b/E_b)_1$. But instead of the constant $\Gamma_1 = 5/3$ and 2, the constant $\Gamma_1 = 4/3$ would be used for the envelope. We would show that although the `compatibility criterion' is satisfied by all stable (as well as unstable) models comprising the sequence but still, the $M-R$ relation does not provide the necessary and sufficient condition of dynamical stability, since the `compatibility criterion' is not `appropriately' satisfied by the equilibrium configurations.

{\bf Step (4):} We would re-construct the NS sequences described in steps (2) and (3) respectively for an `appropriate' boundary value of $(P_b/E_b) = (P_b/E_b)_2$ (say). Since for this particular value of $(P_b/E_b)$, The `compatibility criterion' turns out to be `appropriately' satisfied by all the equilibrium sequences corresponding to an envelope with $\Gamma_1 = 4/3, 5/3$ and 2 respectively, it would follow that the $M-R$ relation  does provide the necessary and sufficient condition of dynamical stability for the said sequences [the value $(P_b/E_b)_2$ corresponds to the {\em minimum} value of $(P_b/E_b)$ for which the $M-R$ relation fulfills the necessary and sufficient condition of dynamical stability for the equilibrium sequence corresponding to NS models with an envelope $\Gamma_1 = 2$. Since the value $(P_b/E_b)_2$ turns out to be greater than $(P_b/E_b)_1$, it follows, therefore,  that this particular value $(P_b/E_b)_2$ would guarantee the automatic fulfillment of the `compatibility criterion' for all the sequences corresponding to NS models with an envelope $\Gamma_1 = 4/3, 5/3$ and 2 respectively, together with fulfilling the necessary and sufficient condition of dynamical stability provided by the $M-R$ relation. This condition is termed as the `appropriate' fulfillment of `compatibility criterion'.].

The methodology regarding the construction of various models mentioned in steps (1) - (4) above and the important outcome emerging thereafter are discussed in the following section.

The  assignment of different values of $\Gamma_1$ between  the 
range (4/3) and 2 in the envelope of NS models discussed under step (2) to step (4) above follows from the fact that  (4/3) 
represents the EOS of ultrarelativistic degenerate electrons  and 
non-relativistic nuclei (Chandrasekhar 1935) or of  relativistic 
degenerate neutron gas (Zeldovich \& Novikov 1978). (5/3) 
corresponds to the well known EOS of non-relativistic  degenerate 
neutrons gas (Oppenheimer \& Volkoff 1939) and $\Gamma_1 = 2$  represents 
the case of  extreme  relativistic  baryons  interacting  through 
vector meson field (Zeldovich 1962) [The value of $\Gamma_1  > 2$ is also  possible 
for some EOS, e.g.,  Malone,  Johnson \&  Bethe  (1975);  Clark, 
Heintzmann \& Grewing (1971), however, the  results  obtained  in 
this paper do not get affected by choosing $\Gamma_1  > 2$]. The  various 
values of $\Gamma_1$ chosen in this range ($4/3 \leq \Gamma_1 \leq 2$) can cover almost all of the  nuclear 
EOS discussed in the literature (and, therefore, it can cover almost full range of nuclear density which might be applicable for the envelope region as well), and in our opinion, it could be more appropriate 
to choose an average (constant) value of $\Gamma_1$ for a conventional NS model,  instead  of  going 
into the details of the  density  range  below $E_b$   specified  by 
different EOS [for example, BPS, NV, or FPS (Lorenz, Ravenhall \& 
Pethick 1993)] which are frequently used by various  authors  in 
the conventional models of  neutron  stars in-spite of their uncertainty (Friedman \& Ipser 1987; Dolan 1992) as mentioned earlier. The EOS of
classical polytrope for the said  $\Gamma_1$ values, considered in the present study, will not only simplify the procedure
but also provide necessary insight regarding the suitability of the EOS for the envelope region, since this study yields the important finding that the stable sequences of neutron  star models terminate at the same value of maximum mass independent of the EOS of the envelope. 


\section{Methodology and Discussion}
The metric for spherically symmetric and static configurations can be written
in the following form (remembering that we are using geometrized units, i.e. $G = c = 1$; where $G$ and $c$ represent respectively, the universal constant of gravitation and the speed of light in vacuum) 
\begin{equation}
ds^2 =  e^{\nu} dt^2 - e^{\lambda} dr^2 - r^2 d\theta^2 - r^2 $ sin$^2 \theta d\phi^2 , 
\end{equation}
where $\nu$ and $\lambda$ are functions of $r$ alone.  The  Oppenheimer-Volkoff 
(O-V) equations (Oppenheimer \& Volkoff 1939), resulting  from  Einstein's field  equations, for 
systems with isotropic pressure $P$  and  energy-density  $E$  can  be 
written as

\begin{figure}
   \centering
\psfig{file=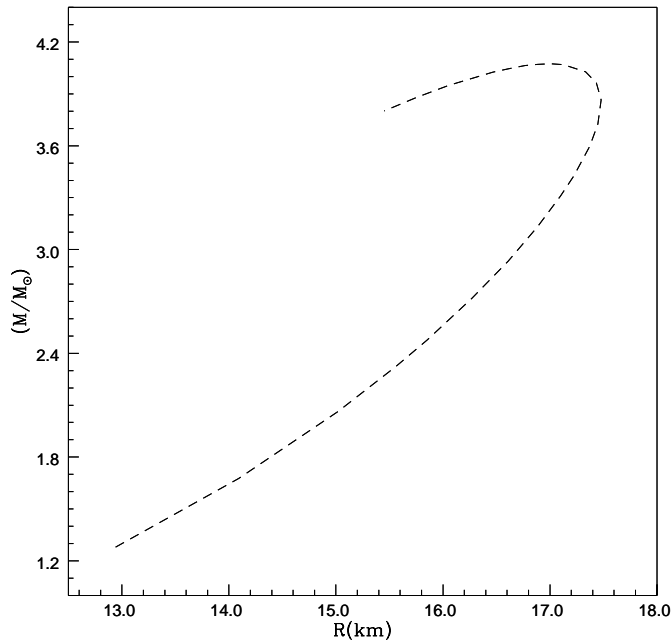,height=9.5cm,width=9.5cm} 
      \caption{The mass-radius diagram of the models corresponding to the stiffest EOS, $P = (E - E_s)$, (as discussed in the text (step 1, sections 2 and 3)) for an assigned value of the density $E_s = 2.7 \times 10^{14}$ g\, cm$^{-3}$ at the surface. The models do not fulfill the `compatibility criterion' as shown in Table 1. Also, the mass-radius diagram does not provide the necessary and sufficient condition of dynamical stability because the inequalities, $(\Gamma_1)_0 \leq 2.5946$ and $u_{\rm max} \leq 0.3406$, are not fulfilled simultaneously at the maximum value of mass as shown in Table 1.
              }
         \label{FigVibStab}
   \end{figure}

\begin{eqnarray}
P' & = & - (P + E)[4 \pi P r^3 + m]/r(r - 2m) \\                       
\nu'/2 & = & - P'/(P + E)  \\                                            
m'(r) & = & 4\pi E r^2 \,;
\end{eqnarray}
where $ m(r) = \int_{0}^{r} 4\pi Er^2 dr $ is the mass, contained within the  radius  $r$,  and  the 
prime denotes radial derivative.

\begin{figure}
\centering
\psfig{file=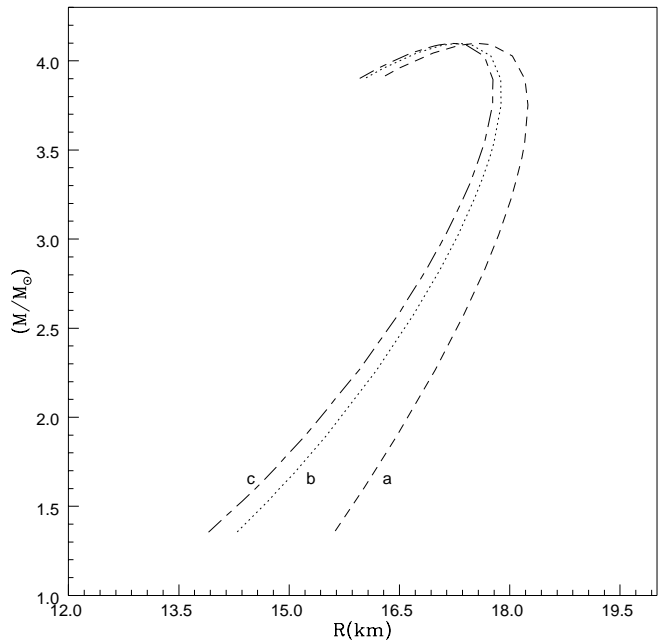,height=9.5cm,width=9.5cm} 
      \caption{The mass-radius diagram of the models as discussed in the text (steps (2) and (3), sections 2 and 3) for an assigned value of matching density
            $E = E_b = 2.7 \times 10^{14}$ g\, cm$^{-3}$ at the core-envelope
            boundary. The labels a, b and c represent the models for an envelope with $\Gamma_1 =  (4/3), 
            (5/3)$, and 2 respectively. The value  of  the  ratio  of 
pressure to energy-density, $(P_b/E_b) = (P_b/E_b)_1 = 1.0645 \times 10^{-2}$, at the core envelope boundary is assigned in such a manner that the `compatibility criterion' has become satisfied by all the models corresponding to an envelope with $\Gamma_1 = 4/3$, whereas it remains unsatisfied by various models corresponding to an envelope with $\Gamma_1 = 5/3$ and 2 respectively as shown in Table 3. Since, the `compatibility criterion' has not been satisfied `appropriately' by the models corresponding to an envelope with $\Gamma_1 = 4/3$ (and has not been satisfied by various models corresponding to an envelope with $\Gamma_1 = 5/3$ and 2), the $M-R$ relation does not provide the necessary and sufficient condition of dynamical stability for any sequence as shown in Table 2.
              }
         \label{FigVibStab}
   \end{figure} 

In order to solve equations (3) - (5) for different models considered in the present study, we consider the stiffest EOS

\begin{figure}
\centering
\psfig{file=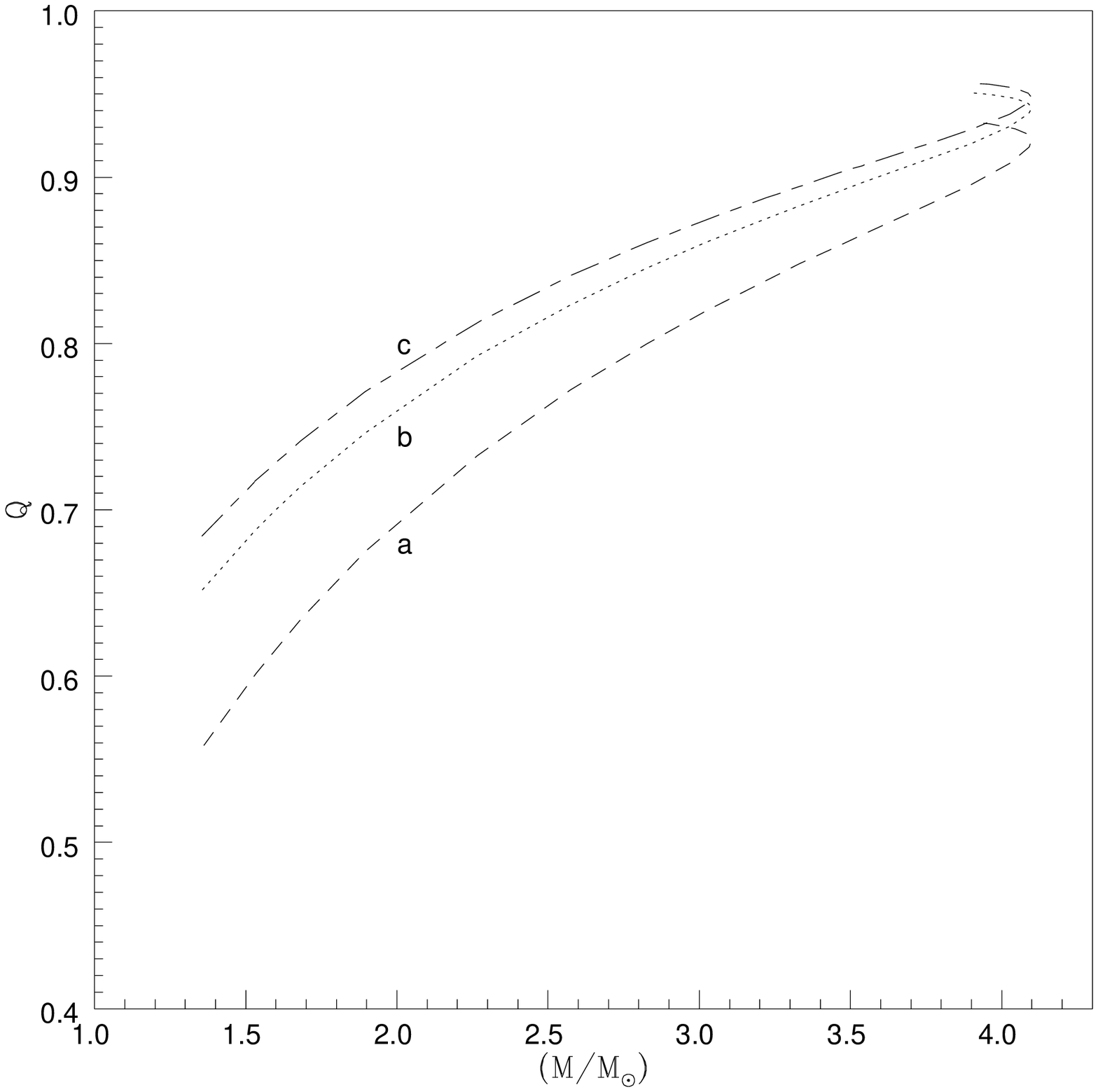,height=9.5cm,width=9.5cm} 
      \caption{Fractional moment of inertia $Q(\equiv I_{\rm core}/I_{\rm total})$ vs. total mass $M$ for the models presented in Tables 2 and 3 and shown in Figure 2. The labels a, b and c denote the models for an envelope with $\Gamma_1 =  (4/3), 
            (5/3)$, and 2 respectively.}
         \label{FigVibStab}
   \end{figure}

\begin{equation}
P = E - E_s,
\end{equation}
for the entire configuration (or for the core region), and the EOS of classical polytrope
\begin{equation}
\frac {{\rm dln}P}{{\rm dln}\rho} = \Gamma_1
\end{equation}
for the envelope region respectively. For core-envelope models, the fractional moment of inertia (given by equation (1)) may be calculated by using an approximate but very precise empirical formula which is based on 30 theoretical EOSs of dense nuclear matter. For NS models, the formula yields in the following form (Bejger \& Haensel 2002)
\begin{equation}
I \simeq \frac{2}{9}(1 + 5x)MR^2, \,\,\,\, x > 0.1 
\end{equation}

where $x$ is the compactness parameter measured in units of $[M_\odot({\rm km})/{\rm km}]$, i.e.
\begin{equation}
x = \frac{M/R}{M_\odot/{\rm km}} = \frac{u}{1.477}.
\end{equation}

In accordance with the four step consecutive study mentioned in the last section, we construct different models in the following manner

{\bf Models mentioned in step (1):} The coupled equations (3 - 5) are solved for  the  EOS given by equation (6) until pressure vanishes at the surface of the configuration for a fiduciary choice of $E_s = 2.7 \times 10^{14}$ g\, cm$^{-3}$, the nuclear saturation density. At the surface, $r = R$, we obtain $P = 0, E = E_s, m(r = R)  =  M, e^{\nu} = e^{-\lambda} = (1 - 2M/R) = (1 - 2u)$. The results of the calculations are shown in Table 1 and the $M-R$ diagram is presented in Fig.1. It follows from Table 1 that along the stable branch of the sequence, the maximum value of mass corresponds to the maximum `stable' value of $u_{\rm max} \simeq 0.3539$ and the corresponding `local' value of $(\Gamma_1)_0 \simeq 2.4990$. Although, this value of $(\Gamma_1)_0$ turns out to be consistent with that of the absolute upper bound on $(\Gamma_1)_0 ((\Gamma_1)_{0,\rm max, abs} \cong 2.5946)$, the maximum `stable' value of $u_{\rm max} \simeq 0.3539$ is found to be inconsistent with that of the absolute upper bound on $u_{\rm max} (u_{\rm max,abs} \cong 0.3406)$. Thus the configuration turns out to be inconsistent with the corollary 1 of Theorem 2 (Negi 2007). It follows, therefore, that the $M-R$ relation corresponding to the stiffest EOS does not provide the necessary and sufficient condition of dynamical stability for the equilibrium configurations. Since, the total mass `$M$' which appears here does not fulfill the definition of the `actual mass' which should be present in the exterior Schwarzschild solution (Negi 2004; 2006), the equilibrium sequence corresponding to the pure stiffest EOS, therefore, does not even fulfill the necessary condition of hydrostatic equilibrium (Negi 2007), as a result the `compatibility criterion' can not be made satisfied by such configurations. This is also evident from the comparison of column 2 and 6 of Table 1 that for each assigned value of $(P_0/E_0),\, u_{\rm stff} > u_h$.

{\bf Models mentioned in step (2):} The coupled equations (3 - 5) are solved by considering equation (6) in the core ($0 \leq r \leq b$) and equation (7) in the envelope ($b \leq r \leq R$) for the constant $\Gamma_1 = 5/3$ and 2 respectively, until the boundary conditions: $P  = E = 0, m(r = R)  =  M, e^{\nu} = e^{-\lambda} = (1 - 2M/R) = (1 - 2u)$ are reached at the surface, $r = R$, of the configuration. Equation (8) is also used together with equations (3 - 5) to calculate the fractional moment of inertia given by equation (1) and the moment of inertia of the entire core-envelope model considered in the present section. The calculations are performed for a fiduciary choice of the boundary density, $E_b  =  2.7 \times 10^{14}$ g\, cm$^{-3}$, and for an `arbitrarily' assigned value of the ratio of pressure to energy-density at the core-envelope boundary, $(P_b/E_b) = (P_b/E_b)_1 = 1.0645 \times 10^{-2}$. The results of the calculations are summarized in Tables 2 - 3 (indicated with a superscript `b' for the $\Gamma_1 = 5/3$ and `c' for the $\Gamma_1 = 2$ envelope models) and the $M-R$ diagram is shown in Fig.2 (label `b' for the $\Gamma_1 = 5/3$ and `c' for the $\Gamma_1 = 2$ envelope models). Though, the construction of such core-envelope models appropriately fulfill the definition of the `actual mass' appearing in the exterior Schwarzschild solution (Negi 2007), the choice of $(P_b/E_b) =(P_b/E_b)_1$, however, yields the sequences (corresponding to NS models with an envelope $\Gamma_1 = 5/3$ and 2 respectively) which fulfill only the necessary (but not sufficient) condition of hydrostatic equilibrium, since the `compatibility criterion' remains unsatisfied by such sequences as shown in Table 3. From the Table 2, we find that at the maximum mass values along the stable branch of the sequences, the sequence composed of NS models with an envelope $\Gamma_1 = 5/3$ yields $u_{\rm max} \simeq 0.3493$  and the corresponding value of $(\Gamma_1)_0 \simeq 2.4984$, whereas the sequence composed of NS models with an envelope $\Gamma_1 = 2$ yields $u_{\rm max} \simeq 0.3509$ and the corresponding value of $(\Gamma_1)_0 \simeq 2.4984$ respectively. Both pairs of these values are, however, found to be inconsistent with that of the pair of absolute upper bounds $u_{\rm max,abs} (\cong 0.3406)$ and $(\Gamma_1)_{0,\rm max,abs} (\cong 2.5946)$ respectively. It follows, therefore, that the $M-R$ relation corresponding to the sequences (composed of NS models with an envelope $\Gamma_1 = 5/3$ and 2 respectively) does not provide the necessary and sufficient condition of dynamical stability as shown in Fig.2.

Tables 2 - 3 and Fig.3 show that the range of fractional moment of inertia, $0.652 \leq Q \leq 0.948$, corresponding to the `stable' models (with an envelope $\Gamma_1 = 5/3$ and 2 respectively) possesses the NS masses in the range $1.35M_\odot \leq M \leq 4.1M_\odot$. This feature is consistent with those of the conventional models discussed in the literature and can explain only the higher values of the glitch healing parameter observed for the Crab-like pulsars. However, for the minimum weighted mean value of $Q \simeq 0.7$ corresponding to the Crab pulsar, we obtain from Fig.3, the minimum masses $M_b \simeq 1.6M_\odot$ and $M_c \simeq 1.44M_\odot$ for the Crab pulsar.

{\bf Models mentioned in step (3):} In order to construct core-envelope models corresponding to the constant $\Gamma_1 =4/3$ for the envelope, we use equation (7) for the envelope ($b \leq r \leq R$) and equation (6) for the core ($0 \leq r \leq b$) and solve equations (3 - 5) together with equation (8) for the same boundary conditions mentioned in the last sub-section (step (2)) for the surface and the core-envelope boundary, viz.  $E_b  =  2.7 \times 10^{14}$ g\, cm$^{-3}$, and $(P_b/E_b) =(P_b/E_b)_1 = 1.0645 \times 10^{-2}$ respectively. The various parameters obtained for this model are indicated with a superscript `a' in Tables 2 and 3 respectively, and the $M-R$ diagram (marked with label `a') is presented in Fig.2. Table 3 shows that although the `compatibility criterion' is satisfied by all members of the sequence corresponding to NS models with an envelope $\Gamma_1 = 4/3$ for the choice of the boundary condition $(P_b/E_b) =(P_b/E_b)_1$, this choice, however, does not appear to be `appropriate' because at the maximum value of mass along the stable branch of the sequence in the mass-radius relation, the model yields the maximum value of $u (u_{\rm max}) \simeq 0.3444$ and the corresponding value of $(\Gamma_1)_0 \simeq 2.4984$ as shown in Table 2. These values, however, show inconsistency with that of the pair of absolute upper bounds, $u_{\rm max,abs}(\cong 0.3406)$ and $(\Gamma_1)_{0,\rm max,abs} (\cong 2.5946)$, respectively. It follows from this result that the necessary and sufficient condition of dynamical stability (provided by the $M-R$ relation) may remain unsatisfied even when all members of the sequence do satisfy the `compatibility criterion'.

Tables 2 - 3 and Fig.3 also indicate that the range of fractional moment of inertia, $0.558 \leq Q \leq 0.922$, for the `stable' sequence corresponding to the $\Gamma_1 = 4/3$ envelope model provides the masses of NS models in the range, $1.36M_\odot \leq M \leq 4.1M_\odot$. These higher values of the glitch healing parameter are consistent only with those the Crab-like pulsars as discussed in the last sub-section (step 2). However, for the minimum value of $Q \simeq 0.7$, Fig.3 yields the minimum mass $M_c \simeq 2.03M_\odot$ for the Crab pulsar.

\begin{figure}
   \centering
\psfig{file=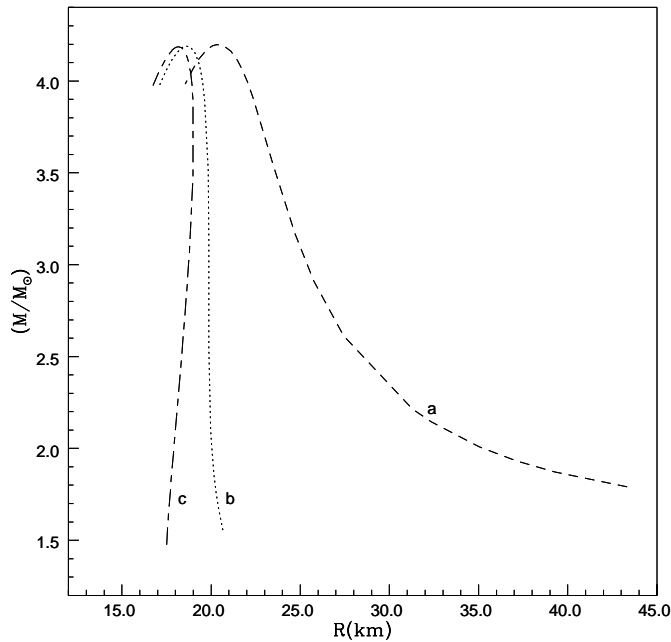,height=9.5cm,width=9.5cm} 
      \caption{The mass-radius diagram of the models as discussed in the text (step 4, sections 2 and 3) for an assigned value of matching density
            $E = E_b = 2.7 \times 10^{14}$ g\, cm$^{-3}$ at the core-envelope
            boundary. The labels a, b and c represent the models for an envelope with $\Gamma_1 =  (4/3), 
            (5/3)$, and 2 respectively. The value  of  the  ratio  of 
pressure to energy-density, $(P_b/E_b) = (P_b/E_b)_2 = 4.694 \times 10^{-2}$, at the core envelope boundary is obtained in such a manner that the `compatibility criterion' is `appropriately' satisfied by {\em all} the models corresponding to an envelope with $\Gamma_1 = 4/3, 5/3$ and 2 respectively, as shown in Table 5. That is, the necessary and sufficient condition of dynamical stability provided by the $M-R$ relation is also satisfied together with satisfying the `compatibility criterion' as shown in Table 4.
              }
         \label{FigVibStab}
   \end{figure}

{\bf Models mentioned in step (4):} We re-construct the NS sequences (composed of core-envelope models), those described in the last two sub-sections (steps (2) and (3) respectively), but for an `appropriate' value of $(P_b/E_b) =(P_b/E_b)_2 = 4.694 \times 10^{-2}$.  The results of the calculations are summarized in Tables 4 - 5 and the $M-R$ diagram is shown in Fig.4. The superscripts `a', `b' and `c' which appear among various parameters in these tables, represent the models with an envelope $\Gamma_1 = 4/3, 5/3$ and 2 respectively. The $M-R$ diagram is also labeled in the similar fashion. The value $(P_b/E_b)_2$, in fact, represents the {\em minimum} value of $(P_b/E_b)$ for which {\em all} the sequences, corresponding to NS models with an envelope $\Gamma_1 = 4/3, 5/3$ and 2 respectively, fulfill the necessary and sufficient condition of dynamical stability provided by the $M-R$ relation. Since the value $(P_b/E_b)_2 > (P_b/E_b)_1$, it follows that the `compatibility criterion' would automatically be satisfied  by {\em all} the sequences corresponding to NS models with an envelope $\Gamma_1 = 4/3, 5/3$ and 2 respectively. This is evident from the Tables 4 and 5 respectively. A comparison of column 2 with those of column 3, 4 and 5 of Table 5 show that the `compatibility criterion' is `appropriately' satisfied by all the sequences. Table 4 shows that at the maximum values of mass along the stable branch of the sequences, the models corresponding to an envelope with $\Gamma_1 = 4/3, 5/3$ and 2 yield the maximum values of $u (u_{\rm mas}) \simeq 0.3044$, 0.3324 and 0.3404 respectively for the same value of $(\Gamma_1)_0 \simeq 2.4955$ . Evidently, these values are consistent  with those of the absolute upper bounds $u_{\rm max,abs}(\cong 0.3406)$ and $(\Gamma_1)_{0,\rm max,abs} (\cong 2.5946)$ respectively.

\begin{figure}
\centering
\psfig{file=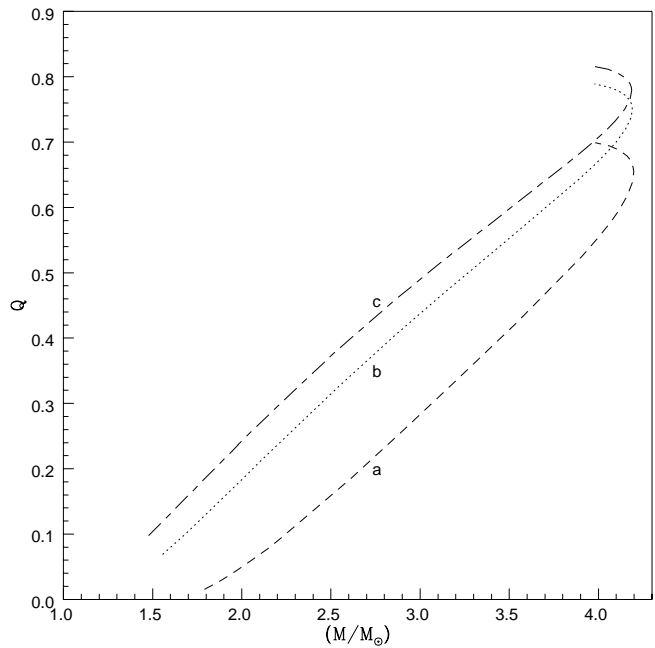,height=9.5cm,width=9.5cm} 
      \caption{Fractional moment of inertia $Q(\equiv I_{\rm core}/I_{\rm total})$ vs. total mass $M$ for the models presented in Tables 4 and 5 and shown in Figure 4. The labels a, b and c denote the models for an envelope with $\Gamma_1 =  (4/3), 
            (5/3)$, and 2 respectively.}
         \label{FigVibStab}
   \end{figure}

Tables 4 - 5 and Fig.5 show that the ranges of fractional moment of inertia, $0.069 \leq Q \leq 0.779$, obtained for the stable sequences corresponding to the $\Gamma_1 = 5/3$ and 2 envelope models have masses in the range, $1.48M_\odot \leq M \leq 4.2M_\odot$. Obviously, these ranges are capable of fulfilling the higher as well as lower values of the glitch healing parameter corresponding to the case of the Crab and the Vela-like pulsars respectively. For the maximum weighted mean value of $Q \simeq 0.2$ corresponding to the Vela pulsar, Fig.5 yields the maximum masses $M_b \simeq 2.06M_\odot$ and $M_c \simeq 1.85M_\odot$ for the Vela pulsar, whereas for the minimum weighted mean value of $Q \simeq 0.7$, the models yield the minimum masses $M_b \simeq 4.1M_\odot$ and $M_c \simeq 3.98M_\odot$ for the Crab pulsar respectively. It also follows from the Tables 4 - 5 and Fig.5 that for the stable range of fractional moment of inertia, $0.016 \leq Q \leq 0.655$, corresponding to the $\Gamma_1 = 4/3$ envelope models, the NSs can have masses in the range, $1.79M_\odot \leq M \leq 4.2M_\odot$. For the maximum weighted mean value of $Q \simeq 0.2$, Fig.5 yields the maximum mass $M_c \simeq 2.67M_\odot$ for the Vela pulsar, however the minimum weighted mean value, $Q \simeq 0.7$, corresponds to the minimum mass, $M_a \approx 3.98M_\odot$, for the Crab pulsar which lies in the unstable branch of Fig.5.

The important conclusions emerge from the four step study and discussion presented above may be summarized in the following manner:

(1) Steps (1), (2) and (4) show that the $M-R$ relation provides the necessary and sufficient condition of dynamical stability only when the `compatibility criterion' for the equilibrium configurations is satisfied. Without satisfying the `compatibility criterion', the $M-R$ relation can not provide the necessary and sufficient condition of dynamical stability.

(2) The inclusion of step (3), however, modifies the above conclusion in the following statement:
The $M-R$ relation provides the necessary and sufficient condition of dynamical stability only when the `compatibility criterion' is `appropriately' satisfied. Thus, the necessary and sufficient condition provided by the $M-R$ relation is totally dependent on the `appropriate' fulfillment of `compatibility criterion'. On the other hand, the fulfillment of `compatibility criterion' is quite independent of the fulfillment of necessary and sufficient condition of dynamical stability provided by the $M-R$ relation, because the `compatibility criterion' can be made satisfied even without fulfilling the necessary and sufficient condition of dynamical stability provided by the $M-R$ relation. Since, the fulfillment of the `compatibility criterion' alone is a measure of the fulfillment of necessary and sufficient condition of hydrostatic equilibrium for any static and spherical configuration (Negi 2007), it follows from the above discussion that the fulfillment of the `compatibility criterion' alone, independently provides, in general, the necessary and sufficient condition of hydrostatic equilibrium for any sequence composed of regular, static and spherical configurations.

\begin{figure}
\centering
\psfig{file=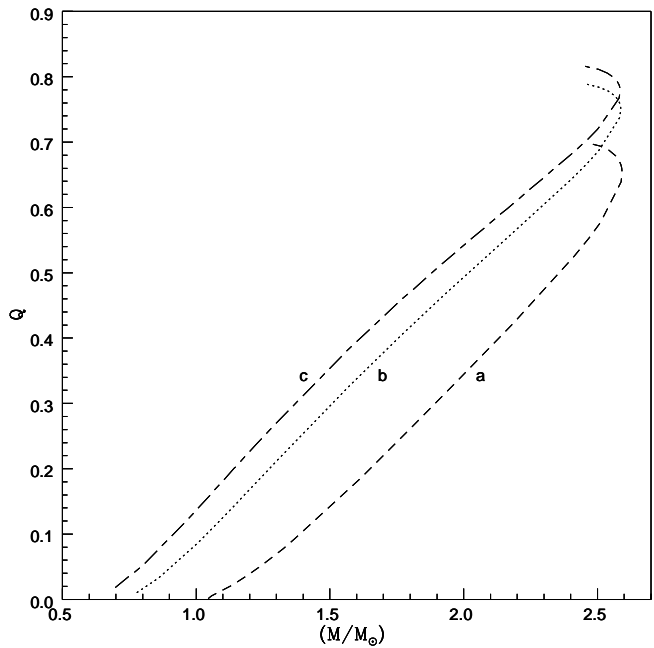,height=9.5cm,width=9.5cm} 
      \caption{Fractional moment of inertia $Q(\equiv I_{\rm core}/I_{\rm total})$ vs. total mass $M$ for the models as discussed in the text (step 4, section 2 and 3) for a {\em calculated} value of matching density
            $E = E_b = 7.0794 \times 10^{14}$ g\, cm$^{-3}$ at the core-envelope
            boundary. The labels a, b and c represent the models for an envelope with $\Gamma_1 =  (4/3), 
            (5/3)$, and 2 respectively. The value  of  the  ratio  of 
pressure to energy-density, $(P_b/E_b) = (P_b/E_b)_2 = 4.694 \times 10^{-2}$, at the core envelope boundary is obtained in such a manner that the `compatibility criterion' is `appropriately' satisfied for {\em all} the models corresponding to an envelope with $\Gamma_1 = 4/3, 5/3$ and 2 respectively (in the same manner as shown in Table 5 for the matching density $E = E_b = 2.7 \times 10^{14}$ g\, cm$^{-3}$ at the core-envelope boundary). That is, the necessary and sufficient condition of dynamical stability provided by the $M-R$ relation is also satisfied together with satisfying the `compatibility criterion' (in the same manner as shown in Table 4 for the matching density $E = E_b = 2.7 \times 10^{14}$ g\, cm$^{-3}$ at the core-envelope boundary).}
         \label{FigVibStab}
   \end{figure}

\section{Results and conclusions}

We have investigated that the mass-radius relation corresponding to the stiffest  equation  of  state does not provide the necessary and sufficient condition of dynamical stability because at the maximum value of mass along the mass-radius relation, the pair of the maximum `stable' value of compactness, $u_{\rm max} \simeq 0.3539$, and the corresponding central value of the `local' adiabatic index, $(\Gamma_1)_0 \simeq 2.4990$, turns out to be inconsistent with that of the pair of {\em absolute} values,
$u_{max,abs} \cong 0.3406$, and $(\Gamma_1)_{0,{\rm max,abs}} \cong 2.5946$, compatible with the structure of general relativity, causality, and
dynamical stability. The reason behind this inconsistency lies in the fact that the `compatibility criterion' (Negi and Durgapal 2001) can not be fulfilled by any of the sequence, composed of regular configurations corresponding to a {\em single} EOS with finite (non-zero) values of surface and central density (Negi 2004; 2006).

We have further  constructed the sequences composed of core-envelope models such that, like the stiffest EOS, each member of these sequences satisfy the extreme case of causality condition, $v = c = 1$, at the centre by considering the stiffest  equation  of  state in  the  core  and  a 
polytropic  equation  with the constant   adiabatic   index  $\Gamma_1 = [$d$lnP/$d$ln\rho]$ in the envelope and investigated the resulting configurations on the basis of the `compatibility criterion' and the mass-radius relation for different values of $\Gamma_1 = 4/3, 5/3$ and 2. Together with the finding corresponding to the case of stiffest EOS mentioned above, the investigation of the said sequences of the core-envelope models explicitly show that the mass-radius relation corresponding to the said sequences can provide the necessary and sufficient condition of dynamical stability only when the `compatibility criterion' for the sequences in `appropriately' satisfied. However, the fulfillment of `compatibility criterion' can remain satisfied even when the mass-radius relation does not provide the necessary and sufficient condition of dynamical stability for the equilibrium configurations.

In continuation to the results of previous study (Negi 2007), these results explicitly show that the `compatibility criterion' {\em independently} provides, in general, the {\em necessary} and {\em sufficient} condition of hydrostatic equilibrium for any regular sequence.

The causal sequences of NS models constructed in the present study (which fulfill the necessary and sufficient condition of hydrostatic equilibrium and dynamical stability simultaneously) terminate at the same value of maximum mass, $M_{max} \approx 4.2 M_\odot$ (for $E_b  = 2.7 \times 10^{14}$ g\, cm$^{-3}$), independent of the EOS of the envelope. However, the maximum compactness, $u_{\max} \simeq 0.3404$, yields for the sequence corresponding to the  $\Gamma_1 = 2$ envelope models. While for the the same value of transition density $E_b  = 2.7 \times 10^{14}$ g\, cm$^{-3}$ at the core-envelope boundary, this upper bound on NS mass  found to be fully consistent with those of the models formulated with the advance nuclear theory (Kalogera \& Baym 1996), the $\Gamma_1 = 2$ envelope model indicates the appropriateness of the (average) value of $\Gamma_1$ for the entire envelope.

Beside its fundamental feature, this study also underlines the importance of the applicability of `compatibility criterion' to the conventional models of NSs. Since, we find that when the `compatibility criterion' is not satisfied (for the case of the models corresponding to an envelope with $\Gamma_1 = 5/3$ and 2, if the ratio of pressure to energy-density at the core-envelope boundary, $P_b/E_b$, is `arbitrarily' assigned to be about $1.0645 \times 10^{-2}$) or not `appropriately' satisfied (for the case of the models corresponding to an envelope with $\Gamma_1 = 4/3$, if $P_b/E_b \simeq 1.0645 \times 10^{-2}$) by the sequences, the corresponding range of the glitch healing parameter turns out to be $0.558 \leq Q \leq 0.948$. This feature is consistent with the other conventional NS models discussed in the literature and can explain only the higher values of $Q$ on the basis of starquake model of glitch generation for the Crab-like pulsars. For the minimum weighted mean value of $Q\simeq 0.7$, our models (corresponding to an envelope with $\Gamma_1 = 2, 5/3$ and 4/3 respectively) yield the minimum masses $M_c \simeq 1.44M_\odot,\,M_b \simeq 1.6M_\odot$ and $M_a\simeq 2.03M_\odot$ for the Crab pulsar. Among these values, the first two values are comparable with those of the minimum values $1.35M_\odot$ and $1.65M_\odot$ obtained by Crawford and Demia\'{n}ski (2003) by using the GWM (Glendenning, Weber, \& Moszkowski 1992) and HKP (Haensel, Kutschera, \& Proszynski 1981) EOSs. Whereas, corresponding to $Q \simeq 0.7$, the other five EOSs of the dense nuclear matter considered by Crawford and Demia\'{n}ski (2003) yield the minimum mass $M < 1M_\odot$ for the Crab pulsar (see, e.g. Crawford and Demia\'{n}ski 2003; and references therein). However for the maximum weighted mean value of $Q \simeq 0.2$ corresponding to the Vela pulsar, our models also yield the unrealistically small mass values for the Vela pulsar together with the other models discussed in the literature (Crawford and Demia\'{n}ski 2003).

On the other hand, as soon as the `compatibility criterion' is `appropriately' satisfied by all the models corresponding to an envelope with $\Gamma_1 = 4/3, 5/3$ and 2 respectively (that is, if the ratio of pressure to energy-density at the core-envelope boundary, $P_b/E_b$, is set (and not `arbitrarily' assigned) to be about $4.694 \times 10^{-2}$ for all the sequences), the corresponding range of the glitch healing parameter turns out to be $0.016 \leq Q \leq 0.779$. This range, however, can explain both (the higher as well as lower) values of $Q$ on the basis of starquake model of glitch generation for the Crab, as well as for the Vela-like pulsars. The sequences corresponding to an envelope with $\Gamma_1 = 5/3$ and 2 yield the maximum masses $M_b \simeq 2.06M_\odot,\, M_c \simeq 1.85M_\odot$ for the Vela ($Q \simeq 0.2$) and the minimum masses $M_b \simeq 4.1M_\odot,\,M_c\simeq 3.98M_\odot$ for the Crab ($Q \simeq 0.7$) pulsar respectively. The sequence corresponding to an envelope with $\Gamma_1 = 4/3$, however, yields the maximum mass $M_a \simeq 2.67M_\odot$ for the Vela pulsar ($Q \simeq 0.2$) but it does not provide the minimum stable mass for the Crab pulsar as soon as the constraint of $Q \geq 0.7$ is imposed. 

The higher mass values mentioned in the last paragraph for the Crab pulsar seem to be unlikely, since none of the observational and/or the theoretical study predict such higher mass values for the Crab pulsar. Thus, unlike the results of steps 2 - 3 (section 3) which deal with the study of NS models without implementing the `appropriate' fulfillment of `compatibility criterion', the implementation of the `appropriate' fulfillment of `compatibility criterion' also reveals that in order to construct a `realistic' NS sequence composed of NS masses comparable with those of the observations, we have to modify the value of matching density at the core-envelope boundary. In view of the modern EOSs of dense nuclear matter, the upper bound on NS mass compatible with causality and dynamical stability can reach up to a value as large as $2.2M_\odot$, since among the variety of modern EOSs discussed in the literature only the following EOSs yield the maximum mass of NS model in excess of 2$M_\odot$: SLy (Douchin \& Haensel 2001) EOS, $M_{\rm max} = 2.05M_\odot$; BGN1 (Balberg \& Gal 1997) EOS, $M_{\rm max} = 2.18M_\odot$; and APR (Akmal et. al. 1998) EOS, $M_{\rm max} = 2.21M_\odot$ (see, e.g. Haensel et al 2006). However, on the basis of other
modern EOSs for NS matter, fitted to experimental 
nucleon-nucleon scattering data and the properties of light nuclei, Kalogera \& Baym (1996; and references therein) have also shown that the lowest possible upper bound on NS mass, compatible with causality and dynamical stability, corresponds to the value of $2.2 M_\odot$ which can exceed up to a value as large as $2.9 M_\odot$. Considering their mean $\sim 2.6M_\odot$ as the most likely recent value to the upper bound on NS masses and substituting this value as an upper bound to our models mentioned in step 4 of section 3, we obtain the `appropriate' value of matching density $E_b = 7.0794 \times 10^{14}$ g\, cm$^{-3}$ at the core-envelope boundary. On the basis of this density, our re-constructed sequences (corresponding to an envelope with $\Gamma_1 = 5/3$ and 2 respectively) yield the maximum masses $M_b \simeq 1.28M_\odot,\, M_c \simeq 1.14M_\odot$ for the Vela ($Q \simeq 0.2$) and the minimum masses $M_b \simeq 2.53M_\odot,\,M_c\simeq 2.45M_\odot$ for the Crab ($Q \simeq 0.7)$ pulsar as shown in Fig.6. The sequence corresponding to an envelope with $\Gamma_1 = 4/3$, however, yields the maximum mass $M_a \simeq 1.66M_\odot$ for the Vela pulsar ($Q \simeq 0.2$) but as shown in Fig.6, the minimum mass of the Crab pulsar ($M_a \approx 2.46M_\odot$) belongs to the unstable branch of the sequence as soon as the constraint of $Q \geq 0.7$ is imposed. Obviously, the other conclusions of the present study, drawn on the basis of assigning the fiduciary density $E_b = 2.7\times 10^{14}$ g\, cm$^{-3}$ at the boundary will remain unaltered.

\section*{Acknowledgments}

The author gratefully acknowledges Prof. M. C. Durgapal for his valuable advice and discussion and the referee for his helpful comments and suggestion that improved the manuscript. The Aryabhatta Research Institute of Observational Sciences (ARIES), Nainital is gratefully acknowledged for providing library and computer-centre facilities.


\begin{table*}
\begin{center}
      \caption[]{Mass ($M$), size ($R$), compactness ratio ($u_{\rm stff} \equiv M/R$)  and the central value of the `local' adiabatic index $(\Gamma_1)_0$ of the  configuration for various values of the ratio of central pressure to central energy-density ($P_0/E_0$) as  obtained  by  assigning a fiduciary value of the density at the surface, $E_s  = 2.7 \times 10^{14}$ g\, cm$^{-3}$ ,  for  the stiffest
EOS, $P = (E - E_s)$. It is seen that for each  assigned  value  of $P_0/E_0$, the inequality, $u_{\rm stff} \leq
u_h$  (where $u_h$  represents the corresponding value of the  compactness ratio 
for the homogeneous density distribution shown in column 2) always remains unsatisfied.  The 
slanted values correspond to the limiting case upto which the configuration 
remains pulsationally `stable'.}


\begin{tabular}{cccccc}

\hline
${(P_0 / E_0)}$ & $u_h$ & $(\Gamma_1)_0$ & $(M/M_{\odot})$    & $R({\rm km})$ & $u_{\rm stff}$ \\
\hline
0.10142 & 0.14343 & 10.8600 & 1.27922 & 12.94001 & 0.14601 \\
0.13324 & 0.17226 & 8.50526 & 1.67675 & 14.09223 & 0.17574 \\
0.16786 & 0.19835 & 6.95734 & 2.06245 & 15.01761 & 0.20284 \\
0.19416 & 0.21528 & 6.15039 & 2.32350 & 15.56089 & 0.22054 \\
0.21115 & 0.22511 & 5.73597 & 2.47752 & 15.85387 & 0.23081 \\
0.24024 & 0.24025 & 5.16250 & 2.71665 & 16.27046 & 0.24661 \\
0.27013 & 0.25389 & 4.70192 & 2.93370 & 16.60906 & 0.26089 \\
0.30125 & 0.26640 & 4.31950 & 3.13171 & 16.88386 & 0.27396 \\
0.33330 & 0.27777 & 4.00030 & 3.30968 & 17.10014 & 0.28587 \\
0.35708 & 0.28536 & 3.80049 & 3.42668 & 17.22415 & 0.29384 \\
0.39796 & 0.29698 & 3.51282 & 3.59800 & 17.37210 & 0.30591 \\
0.43484 & 0.30617 & 3.29970 & 3.72636 & 17.44888 & 0.31543 \\
0.48584 & 0.31722 & 3.05829 & 3.86625 & 17.47830 & 0.32672 \\
0.53423 & 0.32626 & 2.87185 & 3.96327 & 17.43511 & 0.33574 \\
0.58144 & 0.33396 & 2.71987 & 4.02801 & 17.33372 & 0.34322 \\
0.64065 & 0.34236 & 2.56092 & 4.06856 & 17.12608 & 0.35088 \\
0.66710 & 0.34573 & {\sl 2.49903} & {\sl 4.07480} & {\sl 17.00784} & {\sl 0.35386} \\
0.69032 & 0.34852 & 2.44860 & 4.07117 & 16.88677 & 0.35608 \\
0.71136 & 0.35092 & 2.40576 & 4.06364 & 16.76655 & 0.35797 \\
0.73537 & 0.35351 & 2.35986 & 4.04578 & 16.61160 & 0.35973 \\
0.75592 & 0.35563 & 2.32289 & 4.02622 & 16.46719 & 0.36112 \\
0.80562 & 0.36037 & 2.24128 & 3.94943 & 16.05711 & 0.36328 \\
0.83608 & 0.36305 & 2.19606 & 3.88049 & 15.75788 & 0.36372 \\
\hline

\end{tabular}
\end{center}
\end{table*}

\begin{table*}
\begin{center}
      \caption[]{Mass ($M$), size ($R$), compactness ratio ($u\equiv M/R$), and the `local' value of adiabatic index at the centre, $(\Gamma_1)_0$, for different values of the ratio of central pressure to central energy-density, $(P_0/E_0)$, for the core-envelope models discussed in the text (steps (2) and (3), section 3). Various parameters are  obtained  by  assigning a fiduciary value of $E_b  = 2.7 \times 10^{14}$ g\, cm$^{-3}$ for an assigned value of the ratio of pressure to energy-density, $(P_b/E_b) = (P_b/E_b)_1 \simeq 1.0645 \times 10^{-2}$, at the core-envelope boundary. The superscripts a, b, and c which appear among various parameters represent the models corresponding to an envelope with $\Gamma_1 = 4/3, 5/3$, and 2 respectively. The 
slanted values correspond to the limiting case upto which the configurations 
remain pulsationally `stable'. It is seen that the $M-R$ relation does not provide the necessary and sufficient condition of dynamical stability for any of the sequence, since none of the sequences satisfy both of the inequalities, viz. $u_{\rm max} \leq 0.3406$ and $(\Gamma_1)_0\leq 2.5946$, simultaneously at the maximum value of mass. Although, all members of the sequence corresponding to an envelope with $\Gamma_1 = 4/3$ fulfill the `compatibility criterion', $u \leq
u_h$  (where $u_h$  represents the  compactness ratio of homogeneous density sphere for the corresponding value of $P_0/E_0$) without fulfilling the necessary and sufficient condition of dynamical stability provided by the $M-R$ relation, various members in the stable (as well as unstable) branch of the sequences corresponding to an envelope with $\Gamma_1 = 5/3$ and 2 do not fulfill the `compatibility criterion' together with the last condition provided by the $M-R$ relation as shown in Table 3. Thus, the assigned value of $(P_b/E_b) = (P_b/E_b)_1 \simeq 1.0645 \times 10^{-2}$ corresponds to the {\em minimum} value for which the `compatibility criterion', in fact, is not `appropriately' satisfied by the sequence corresponding to an envelope with $\Gamma_1 = 4/3$ as discussed in step (3) of section 3.}

\begin{tabular}{ccccccccccc}

\hline
${(P_0 / E_0)}$ & $(\Gamma_1)_0$  & $(M^a/M_{\odot})$    & $R^a({\rm km})$ & $u^a$ & $(M^b/M_{\odot})$  & $R^b({\rm km})$   & $u^b$ & $(M^c/M_{\odot})$   & $R^c({\rm km})$ & $u^c$ \\ 
\hline
0.10603 & 10.4316 & 1.36098 & 15.62864 & 0.12862 & 1.35663 & 14.29216 & 0.14020 & 1.35439 & 13.90263 & 0.14389 \\
0.11977 & 9.34904 & 1.53480 & 15.90839 & 0.14250 & 1.53103 & 14.71551 & 0.15367 & 1.52908 & 14.36231 & 0.15725 \\
0.12544 & 8.97211 & 1.60425 & 16.02166 & 0.14789 & 1.60069 & 14.87248 & 0.15897 & 1.59885 & 14.54034 & 0.16241 \\
0.13235 & 8.55591 & 1.68726 & 16.14566 & 0.15435 & 1.68392 & 15.06058 & 0.16514 & 1.68218 & 14.73637 & 0.16860 \\
0.15071 & 7.63515 & 1.89840 & 16.46722 & 0.17027 & 1.89555 & 15.50498 & 0.18057 & 1.89405 & 15.21041 & 0.18390 \\
0.18604 & 6.37510 & 2.26582 & 16.98853 & 0.19699 & 2.26364 & 16.18761 & 0.20654 & 2.26249 & 15.94411 & 0.20959 \\
0.22012 & 5.54289 & 2.57456 & 17.37250 & 0.21889 & 2.57282 & 16.68887 & 0.22770 & 2.57189 & 16.47563 & 0.23056 \\
0.25219 & 4.96523 & 2.82777 & 17.65613 & 0.23655 & 2.82633 & 17.04811 & 0.24486 & 2.82554 & 16.85773 & 0.24756 \\
0.28047 & 4.56547 & 3.02420 & 17.84883 & 0.25025 & 3.02296 & 17.29460 & 0.25817 & 3.02230 & 17.12592 & 0.26065 \\
0.31322 & 4.19268 & 3.22347 & 18.01620 & 0.26427 & 3.22240 & 17.51592 & 0.27172 & 3.22181 & 17.35701 & 0.27416 \\
0.33333 & 4.00000 & 3.33234 & 18.08938 & 0.27209 & 3.33138 & 17.61968 & 0.27926 & 3.33085 & 17.47471 & 0.28153 \\
0.35897 & 3.78575 & 3.45739 & 18.16419 & 0.28113 & 3.45653 & 17.72636 & 0.28801 & 3.45603 & 17.58420 & 0.29029 \\
0.37791 & 3.64616 & 3.54070 & 18.20241 & 0.28730 & 3.53990 & 17.78229 & 0.29402 & 3.53946 & 17.65346 & 0.29613 \\
0.43583 & 3.29448 & 3.75240 & 18.24864 & 0.30371 & 3.75176 & 17.88056 & 0.30991 & 3.75139 & 17.76321 & 0.31193 \\
0.48898 & 3.04509 & 3.89663 & 18.20641 & 0.31612 & 3.89610 & 17.87641 & 0.32191 & 3.89580 & 17.76979 & 0.32381 \\
0.56150 & 2.78094 & 4.02745 & 18.03557 & 0.32982 & 4.02703 & 17.74399 & 0.33521 & 4.02677 & 17.64993 & 0.33697 \\
0.63427 & 2.57662 & 4.09052 & 17.74603 & 0.34045 & 4.09018 & 17.48404 & 0.34553 & 4.08998 & 17.40243 & 0.34713 \\
0.66738 & {\sl 2.49839} & {\sl 4.09795} & {\sl 17.57442} & {\sl 0.34440} & {\sl 4.09764} & {\sl 17.32866} & {\sl 0.34926} & {\sl 4.09745} & {\sl 17.24674} & {\sl 0.35090} \\
0.68401 & 2.46197 & 4.09650 & 17.48041 & 0.34613 & 4.09620 & 17.23688 & 0.35100 & 4.09604 & 17.16103 & 0.35253 \\
0.70503 & 2.41839 & 4.08963 & 17.34875 & 0.34817 & 4.08935 & 17.11346 & 0.35294 & 4.08917 & 17.03632 & 0.35452 \\
0.71465 & 2.39928 & 4.08462 & 17.28612 & 0.34901 & 4.08434 & 17.05091 & 0.35380 & 4.08419 & 16.97883 & 0.35529 \\
0.75963 & 2.31644 & 4.04483 & 16.95813 & 0.35229 & 4.04458 & 16.73528 & 0.35696 & 4.04443 & 16.66216 & 0.35851 \\
0.81199 & 2.23154 & 3.95964 & 16.49323 & 0.35459 & 3.95942 & 16.28068 & 0.35920 & 3.95929 & 16.21292 & 0.36069 \\
0.83724 & 2.19440 & 3.90077 & 16.23405 & 0.35490 & 3.90056 & 16.02483 & 0.35951 & 3.90044 & 15.95891 & 0.36099 \\
\hline
\end{tabular}
\end{center}
\end{table*}

\newpage

\begin{table*}
\begin{center}
      \caption[]{Compactness ratio, $u (\equiv M/R)$, and the fractional moment of inertia, $Q(\equiv I_{\rm core}/I_{\rm total)}$, for different values of the ratio of pressure to energy-density, $(P_0/E_0)$, at the centre for the  models presented in Table 2. For each value of $(P_0/E_0)$, the compactness ratio of homogeneous density sphere, $u_h$, is shown in column 2. Various parameters are obtained by assigning a fiduciary value of $E_b  = 2.7 \times 10^{14}$ g\, cm$^{-3}$ for an assigned value of the ratio of pressure to energy-density, $(P_b/E_b) = (P_b/E_b)_1 \simeq 1.0645 \times 10^{-2}$, at the core-envelope boundary. The superscripts a, b, and c which appear among various parameters represent the models corresponding to an envelope with $\Gamma_1 = 4/3, 5/3$, and 2 respectively. The 
slanted values correspond to the limiting case upto which the configurations 
remain pulsationally `stable'. It is seen that for  an  assigned  value  of 
$(P_0/E_0)$ the inequality, $u \leq
u_h$,  is always satisfied (but not `appropriately' satisfied, cf. the caption of Table 2) by all members of the sequence  corresponding to an envelope with $\Gamma_1 = 4/3$, whereas it remains unsatisfied by various members in the stable (as well as unstable) branch of the sequences corresponding to an envelope with $\Gamma_1 = 5/3$ and 2 respectively. It follows from this table that the choice of $P_b/E_b (\simeq 1.0645 \times 10^{-2}$; in the present context, for example) can provide a suitable explanation only for the higher values of the glitch healing parameter $Q$ in the range, $0.558 \leq Q \leq 0.948$ (in the present context, for example), on the basis of the starquake mechanism of glitch generation as shown in Fig.3.}

\begin{tabular}{cccccccc}

\hline
${(P_0 / E_0)}$ & $u_h$ & $u^a$ & $u^b$ & $u^c$ & $Q^a$ & $Q^b$ & $Q^c$ \\ 
 
\hline
0.10603 & 0.14794 & 0.12862 & 0.14020 & 0.14389 & 0.55820 & 0.65183 & 0.68421 \\
0.11977 & 0.16070 & 0.14250 & 0.15367 & 0.15725 & 0.60171 & 0.68740 & 0.71684 \\
0.12544 & 0.16567 & 0.14789 & 0.15897 & 0.16241 & 0.61720 & 0.70036 & 0.72805  \\
0.13235 & 0.17152 & 0.15435 & 0.16514 & 0.16860 & 0.63523 & 0.71436 & 0.74134  \\
0.15071 & 0.18603 & 0.17027 & 0.18057 & 0.18390 & 0.67538 & 0.74645 & 0.77083  \\
0.18604 & 0.21029 & 0.19699 & 0.20654 & 0.20959 & 0.73256 & 0.79226 & 0.81213  \\
0.22012 & 0.23000 & 0.21889 & 0.22770 & 0.23056 & 0.77223 & 0.82325 & 0.84041  \\
0.25219 & 0.24591 & 0.23655 & 0.24486 & 0.24756 & 0.80003 & 0.84534 & 0.86049 \\
0.28047 & 0.25823 & 0.25025 & 0.25817 & 0.26065 & 0.81961 & 0.86085 & 0.87416 \\
0.31322 & 0.27081 & 0.26427 & 0.27172 & 0.27416 & 0.83806 & 0.87524 & 0.88769 \\
0.33333 & 0.27778 & 0.27209 & 0.27926 & 0.28153 & 0.84795 & 0.88286 & 0.89418 \\
0.35897 & 0.28593 & 0.28113 & 0.28801 & 0.29029 & 0.85858 & 0.89113 & 0.90219 \\
0.37791 & 0.29149 & 0.28730 & 0.29402 & 0.29613 & 0.86561 & 0.89685 & 0.90686 \\
0.43583 & 0.30640 & 0.30371 & 0.30991 & 0.31193 & 0.88305 & 0.91052 & 0.91962 \\
0.48898 & 0.31785 & 0.31612 & 0.32191 & 0.32381 & 0.89536 & 0.92014 & 0.92842 \\
0.56150 & 0.33083 & 0.32982 & 0.33521 & 0.33697 & 0.90837 & 0.93055 & 0.93793 \\
0.63427 & 0.34152 & 0.34045 & 0.34553 & 0.34713 & 0.91811 & 0.93842 & 0.94493 \\
0.66738 & 0.34577 & {\sl 0.34440} & {\sl 0.34926} & {\sl 0.35090} & {\sl 0.92180} & {\sl 0.94106} & {\sl 0.94765} \\
0.68401 & 0.34778 & 0.34613 & 0.35100 & 0.35253 & 0.92331 & 0.94251 & 0.94864 \\
0.70503 & 0.35021 & 0.34817 & 0.35294 & 0.35452 & 0.92529 & 0.94399 & 0.95028 \\
0.71465 & 0.35128 & 0.34901 & 0.35380 & 0.35529 & 0.92604 & 0.94480 & 0.95070 \\
0.75963 & 0.35600 & 0.35229 & 0.35696 & 0.35851 & 0.92925 & 0.94739 & 0.95349 \\
0.81199 & 0.36095 & 0.35459 & 0.35920 & 0.36069 & 0.93215 & 0.94997 & 0.95578 \\
0.83724 & 0.36314 & 0.35490 & 0.35951 & 0.36099 & 0.93292 & 0.95075 & 0.95650 \\

\hline

\end{tabular}

\end{center}
\end{table*}


\begin{table*}
\begin{center}
      \caption[]{Mass ($M$), size ($R$), compactness ratio ($u\equiv M/R$), and the `local' value of adiabatic index at the centre, $(\Gamma_1)_0$, for different values of the ratio of central pressure to central energy-density, $(P_0/E_0)$, for the core-envelope models discussed in the text (step (4), section 3). Various parameters are  obtained  by  assigning a fiduciary value of $E_b  = 2.7 \times 10^{14}$ g\, cm$^{-3}$ and for an `appropriate' value of the ratio of pressure to energy-density, $(P_b/E_b) = (P_b/E_b)_2 \simeq 4.694 \times 10^{-2}$, at the core-envelope boundary. The superscripts a, b, and c which appear among various parameters represent the models corresponding to an envelope with $\Gamma_1 = 4/3, 5/3$, and 2 respectively. The 
slanted values correspond to the limiting case upto which the configurations 
remain pulsationally stable. It is seen that the $M-R$ relation does provide the necessary and sufficient condition of dynamical stability for all of the sequences, since all of the sequences satisfy both of the inequalities, viz. $u_{\rm max} \leq 0.3406$ and $(\Gamma_1)_0\leq 2.5946$, simultaneously at the maximum value of mass. Because all members of the sequences corresponding to an envelope with $\Gamma_1 = 4/3, 5/3$ and 2 fulfill the `compatibility criterion', $u \leq
u_h$  (where $u_h$  represents the  compactness ratio of homogeneous density sphere for the corresponding value of $P_0/E_0$), together with fulfilling the necessary and sufficient condition of dynamical stability provided by the $M-R$ relation, it follows, therefore, that the assigned value of $(P_b/E_b) = (P_b/E_b)_2 \simeq 4.694 \times 10^{-2}$ corresponds to the {\em minimum} value for which the `compatibility criterion' is `appropriately' satisfied by all the sequences corresponding to NS models with an envelope $\Gamma_1 = 4/3, 5/3$ and 2 respectively as discussed in step (4) of section 3.}

\begin{tabular}{ccccccccccc}

\hline
${(P_0 / E_0)}$ & $(\Gamma_1)_0$  & $(M^a/M_{\odot})$    & $R^a({\rm km})$ & $u^a$ & $(M^b/M_{\odot})$  & $R^b({\rm km})$   & $u^b$ & $(M^c/M_{\odot})$   & $R^c({\rm km})$ & $u^c$ \\ 
\hline
0.10052 & 10.9487 & 1.79122 & 43.32919 & 0.06106 & 1.55535 & 20.66582 & 0.11116 & 1.47601 & 17.50211 & 0.12456 \\
0.11236 & 9.90023 & 1.87195  & 39.26268 & 0.07042 & 1.67909  &  20.40422 & 0.12154  & 1.61142 &  17.58182  & 0.13537 \\
0.12029 & 9.31353 & 1.93294 & 37.11014 & 0.07693  & 1.76262  & 20.27094  & 0.12843 &  1.70131 & 17.65121 & 0.14236 \\
0.12955 & 8.71910 & 2.00865 & 35.04138 & 0.084665  &  1.85989 & 20.15492  & 0.13630 &  1.80486 & 17.73130 & 0.15034 \\
0.14549 & 7.87332 & 2.14512 & 32.33658 & 0.09798  & 2.02462 & 20.01554  & 0.14940 &  1.97821 & 17.88177 & 0.16340 \\
0.15354 & 7.51305 & 2.21518 & 31.27768 & 0.10461 & 2.10582 & 19.97345 & 0.15572 & 2.06296 & 17.95627 & 0.16969 \\
0.20061 & 5.98487 & 2.61355 & 27.42286 & 0.14077  & 2.54549 & 19.87280  & 0.18919 &  2.51677 & 18.35244 & 0.20255 \\
0.23990 & 5.16877 & 2.91143 & 25.75275 & 0.16698  &  2.86158 & 19.87300  & 0.21268 &  2.83972 & 18.61125 & 0.22536 \\
0.27901 & 4.58413 & 3.17036 & 24.69328 & 0.18963  & 3.13199 & 19.87624 & 0.23274  & 3.11471 & 18.80100 & 0.24469 \\
0.34816 & 3.87220 & 3.54016 & 23.49470 & 0.22255  & 3.51403 &  19.84340  & 0.26156 &  3.50187 & 18.97877 & 0.27253 \\
0.44840 & 3.23013 & 3.90788 & 22.35809 & 0.25816  & 3.89112 &  19.64414 & 0.29256  &  3.88312 & 18.97775 & 0.30222 \\
0.48306 & 3.07012 & 3.99636 & 22.02864 & 0.26795  & 3.98169 & 19.54181 & 0.30094 &  3.97464 & 18.92270 & 0.31024 \\
0.51773 & 2.93149 & 4.06762 & 21.71508 & 0.27667  & 4.05467 & 19.41360 & 0.30848 &  4.04842 & 18.83572 & 0.31746 \\
0.54455 & 2.83637 & 4.11157 & 21.47901 & 0.28273 & 4.09975 & 19.30104 & 0.31373 & 4.09402 & 18.74922 & 0.32251 \\
0.58709 & 2.70330 & 4.16243 & 21.10649 & 0.29128 & 4.15213 & 19.09782 & 0.32112 & 4.14711 & 18.58635 & 0.32956 \\
0.63150 & 2.58354 & 4.19170 & 20.70981 & 0.29895  & 4.18272 & 18.84735  & 0.32778 &  4.17831 & 18.37295 & 0.33589 \\
0.66866 & {\sl 2.49552} & {\sl 4.19788} & {\sl 20.36982} & {\sl 0.30438}  & {\sl 4.18981} &  {\sl 18.61628} & {\sl 0.33242}  &  {\sl 4.18584} & {\sl 18.16228} & {\sl 0.34040} \\
0.68296 & 2.46421 & 4.19565 & 20.23663 & 0.30623  & 4.18789 & 18.51726  & 0.33404 & 4.18408 & 18.07316 & 0.34194 \\
0.69900 & 2.43062 & 4.19032 & 20.08121 & 0.30820 &  4.18289 & 18.40185   & 0.33573 & 4.17923 & 17.96912 & 0.34352 \\
0.72302 & 2.38309 & 4.17593 & 19.84795  & 0.31076 &  4.16896 & 18.22159   & 0.33793 & 4.16552 & 17.79888 & 0.34567 \\
0.74010 & 2.35117 & 4.16090 & 19.67543  & 0.312351 &  4.15423 & 18.08144   & 0.33934 & 4.15094 & 17.66827 & 0.34700 \\
0.77503 & 2.29027 & 4.11752 & 19.30611  & 0.31501 &  4.11141 & 17.77494  & 0.34164 & 4.10839 & 17.38013 & 0.34914 \\
0.80605 & 2.24061 & 4.06269 & 18.95854  & 0.31651 &  4.05700 & 17.47310  & 0.34294 & 4.05419 & 17.08740 & 0.35044 \\
0.83972 & 2.19088 & 3.98260 & 18.54954  & 0.31711 &  3.97730 & 17.10352  & 0.34347 & 3.97469 & 16.72970  & 0.35091 \\
\hline

\end{tabular}

\end{center}
\end{table*}

\newpage

\begin{table*}
\begin{center}
      \caption[]{Compactness ratio, $u (\equiv M/R)$, and the fractional moment of inertia, $Q(\equiv I_{\rm core}/I_{\rm total)}$, for different values of the ratio of pressure to energy-density, $(P_0/E_0)$, at the centre for the  models presented in Table 4. For each value of $(P_0/E_0)$, the compactness ratio of homogeneous density sphere, $u_h$, is shown in column 2. Various parameters are obtained by assigning a fiduciary value of $E_b  = 2.7 \times 10^{14}$ g\, cm$^{-3}$ and for an `appropriate' value of the ratio of pressure to energy-density, $(P_b/E_b) = (P_b/E_b)_2 \simeq 4.694 \times 10^{-2}$, at the core-envelope boundary. The superscripts a, b, and c which appear among various parameters represent the models corresponding to an envelope with $\Gamma_1 = 4/3, 5/3$, and 2 respectively. The 
slanted values correspond to the limiting case upto which the configurations 
remain pulsationally stable. It is seen that for  an  assigned  value  of 
$(P_0/E_0)$, the inequality, $u \leq
u_h$,  is `appropriately' satisfied by all members of the sequences corresponding to an envelope with $\Gamma_1 = 4/3, 5/3$ and 2 respectively. It follows from this table that the `appropriate' choice of $P_b/E_b (\simeq 4.694 \times 10^{-2}$; in the present context, for example) can provide a suitable explanation for both (the higher as well as lower) values of the glitch healing parameter $Q$ in the range, $0.016 \leq Q \leq 0.779$ (in the present context, for example), on the basis of the starquake mechanism of glitch generation as shown in Fig.5.}
\begin{tabular}{cccccccc}

\hline
${(P_0 / E_0)}$ & $u_h$ & $u^a$ & $u^b$ & $u^c$ & $Q^a$ & $Q^b$ & $Q^c$ \\ 
 
\hline
0.10052 & 0.14253 & 0.06106 & 0.11116 & 0.12456 & 0.01548 & 0.06873 & 0.09775 \\
0.11236 & 0.15394 & 0.07042 & 0.12154 & 0.13537 & 0.02739 & 0.09920 & 0.13475 \\
0.12029 & 0.16116 & 0.07693 & 0.12843 & 0.14236 & 0.03739 & 0.12071 & 0.15969  \\
0.12955 & 0.16918 & 0.08466 & 0.13630 & 0.15034 & 0.05089 & 0.14626 & 0.18860  \\
0.14549 & 0.18205 & 0.09798 & 0.14940 & 0.16340 & 0.07777 & 0.19020 & 0.23645  \\
0.15354 & 0.18814 & 0.10461 & 0.15572 & 0.16969 & 0.09261 & 0.21182 & 0.25950  \\
0.20061 & 0.21911 & 0.14077 & 0.18919 & 0.20255 & 0.18555 & 0.32652 & 0.37684  \\
0.23990 & 0.24008 & 0.16698 & 0.21268 & 0.22536 & 0.26028 & 0.40470 & 0.45366 \\
0.27901 & 0.25763 & 0.18963 & 0.23274 & 0.24469 & 0.32671 & 0.46877 & 0.51517 \\
0.34816 & 0.28259 & 0.22255 & 0.26156 & 0.27253 & 0.42326 & 0.55590 & 0.59803 \\
0.44840 & 0.30929 & 0.25816 & 0.29256 & 0.30222 & 0.52531 & 0.64342 & 0.67967 \\
0.48306 & 0.31666 & 0.26795 & 0.30094 & 0.31024 & 0.55283 & 0.66607 & 0.70070 \\
0.51773 & 0.32332 & 0.27667 & 0.30848 & 0.31746 & 0.57715 & 0.68625 & 0.71944 \\
0.54455 & 0.32803 & 0.28273 & 0.31373 & 0.32251 & 0.59401 & 0.70020 & 0.73250 \\
0.58709 & 0.33482 & 0.29128 & 0.32112 & 0.32956 & 0.61777 & 0.71982 & 0.75062 \\
0.63150 & 0.34115 & 0.29895 & 0.32778 & 0.33589 & 0.63925 & 0.73770 & 0.76712 \\
0.66866 & 0.34593 & {\sl 0.30438} & {\sl 0.33242} & {\sl 0.34040} & {\sl 0.65474} &  {\sl 0.75034} & {\sl 0.77916} \\
0.68296 & 0.34766 & 0.30623 & 0.33404 & 0.34194 & 0.66008 & 0.75491 & 0.78336 \\
0.69900 & 0.34952 & 0.30820 & 0.33573 & 0.34352 & 0.66592 & 0.75976 & 0.78778 \\
0.72302 & 0.35220 & 0.31076 & 0.33793 & 0.34567 & 0.67364 & 0.76625 & 0.79403 \\
0.74010 & 0.35401 & 0.31235 & 0.33934 & 0.34700 & 0.67868 & 0.77068 & 0.79815 \\
0.77503 & 0.35751 & 0.31501 & 0.34164 & 0.34914 & 0.68772 & 0.77855 & 0.80543 \\
0.80605 & 0.36041 & 0.31651 & 0.34294 & 0.35044 & 0.69398 & 0.78427 & 0.81111 \\
0.83972 & 0.36336 & 0.31711 & 0.34347 & 0.35091 & 0.69887 & 0.78917 & 0.81587 \\

\hline

\end{tabular}

\end{center}
\end{table*}


\begin{thebibliography}{99}

\bibitem{} Akmal, A., Pandharipande, V. R., \& Ravenhall, D. G., 1998, Phys. Rev. C 58, 1804 (APR)
\bibitem{} Arnett W. D., Bowers R. L., 1977, 
     ApJ Suppl. 33, 415

\bibitem{} Balberg, S., \& Gal A., 1997, Nucl. Phys. A 625, 435 (BGN1)
\bibitem{} Baym, G., Pethick, C., \& Sutherland, P. 1971, ApJ, 170, 299 (BPS)

\bibitem{} Bejger, M., \& Haensel, P. 2002, A\&A 396, 917



   \bibitem{} Brecher, K., \& Caporaso, G., 1976,
      Nature 259, 377

\bibitem{} Chandrasekhar, S. 1935, MNRAS, 95, 207

\bibitem{} Clark, J. W. Heintzmann, H., \& Grewing, M. 1971, Astrophys. Letters 
   10, 21
\bibitem{} Crawford, F., \& Demia\'{n}ski, M. 2003, ApJ 595, 1052

   \bibitem{} Dolan, J. F. 1992, 
      ApJ 384, 249
\bibitem{} Douchin, F., \& Haensel, P. 2001, A\&A 380, 151 (SLy)
   \bibitem{} Friedman, J. L., \& Ipser, J. R. 1987, 
      ApJ 314, 594
\bibitem{} Glendenning, N. K., Weber, F., \& Moszkowski, S. A. 1992, 
      Phys. Rev.C 45, 844 (GWM)
\bibitem{} Haensel, P., Kutschera, M., \& Proszynski, M. 1981, A\&A, 102, 
           299 (HKP)
\bibitem{} Haensel, P., Potekhin, A. Y., \& Yakovlev, D. G. 2006, 
           Neutron Stars 1 - Equation of State and Structure, Springer

    \bibitem{} Haensel, P., \& Zdunik J. L. 1989, 
      Nature 340, 617

    \bibitem{} Hartle, J. B. 1978, 
      Phys. Rep. 46, 201 

\bibitem{} Kalogera, V. \& Baym, G. 1996, ApJ 470, L61

\bibitem{} Lee, T. D. 1975, 
      Rev. Mod. Phys. 47, 267

 \bibitem{} Lorenz, C. P., Ravenhall, D. G., \& Pethick, C. J. 1993, Phys. Rev.   Lett. 70, 379 (FPS)

\bibitem{} Malone, R. C., Johnson, M. B., \& Bethe, H. A. 1975, ApJ 199, 741


\bibitem{} Negi, P. S. 2004, Mod. Phys. Lett. A 19, 2941 (astro-ph/0210018)
\bibitem{} Negi, P. S. 2006, Int. J. Theoretical Phys., 45, 1684 (gr-qc/0401024)

\bibitem{} Negi, P. S. 2007, Int. J. Mod. Phys. D. 16, 35 (astro-ph/0606022)

  \bibitem{} Negi, P. S., \& Durgapal, M. C. 2000,
      A\&A 353, 641

\bibitem{} Negi, P. S., \& Durgapal, M. C. 2001,
      Grav. \& Cosmol. 7, 37 (astro-ph/0312516)
   
    \bibitem{} Negele, J., \& Vautherin, D. 1973, Nucl. Phys., A207, 298 (NV)

\bibitem{} Oppenheimer, J. R., \& Volkoff G. M. 1939,
      Phys. Rev. 55, 374
    
    \bibitem{} Rhoades, C. E. Jr., \& Ruffini, R. 1974,
      Phys. Rev. Lett. 32, 324

 
\bibitem{} Tooper, R. F. 1965, ApJ, 142, 1541

\bibitem{} Zeldovich, Ya. B. 1962, Soviet Phys., JETP, 15, 1158

    \bibitem{} Zeldovich, Ya. B., \& Novikov I. D. 1978, 
      Relativistic  Astrophysics, Vol.1,
      Chicago Univ. Press, Chicago
      (Midway Reprint)



 
\end{thebibliography}
\end{document}